\documentclass[11pt]{article}

\usepackage[a4paper,margin=1in]{geometry}
\usepackage{amsmath,amssymb,amsfonts,bm}
\usepackage{mathtools}
\usepackage{physics}
\usepackage{graphicx}
\usepackage{placeins}
\usepackage{booktabs}
\usepackage{microtype}
\usepackage{hyperref}
\usepackage{authblk}
\usepackage{orcidlink}

\title{Bloch states of an infinite alternating-charge lattice\\
under a transverse electric field}

\date{}
\author[1]{Anand Aruna Kumar\,\orcidlink{0000-0001-6148-2777}}

\affil[1]{Research Engineer, IBM Research, Albany, NY, USA\par
	\href{mailto:anand.aruna.kumar@ibm.com}{anand.aruna.kumar@ibm.com}}
\begin{document}

\maketitle

\begin{abstract}
We consider an infinite one-dimensional array of alternating charges embedded in a two-dimensional configuration space and subjected to a weak static electric field transverse to the lattice axis. Expansion of the Coulomb interaction about a lattice site gives a transverse restoring stiffness \(\kappa\propto\zeta(3)/L^3\), defining the oscillator length \(a_y\), while Bloch periodicity is retained along the lattice direction and the transverse motion is represented in a harmonic basis. The Coulomb matrix elements are governed by the dimensionless scale \(\eta_{r-s}\), which couples the reciprocal-lattice transfer to the transverse oscillator length. A regularized one-dimensional extension is also introduced through Coulomb regularization, yielding a finite diagonal offset and an exponentially decaying even-transfer coupling in reciprocal space. Small \(\eta_{r-s}\) yields logarithmic, parity-dependent transverse-state coupling, while large \(\eta_{r-s}\) approaches the one-dimensional Coulomb limit. In the finite-order spectrum, the latter approaches the Kronig–Penney reference toward the band edge while retaining distinct Brillouin-zone curvature. The formulation therefore identifies two characteristic quantities of the alternating lattice: the transverse restoring stiffness and the one-dimensional Madelung form, while retaining the two-dimensional structure required to describe the response to a transverse electric field.
\vspace{0.5pc}

\end{abstract}

\section{Introduction}
\label{sec:sec1}
We consider an infinite array of alternating charges placed along the $x$ axis at
\begin{equation}
    x_n=nL,
    \qquad
    q_n=(-1)^n q,
    \qquad
    n\in\mathbb{Z},
\end{equation}
where $L$ is the nearest-neighbor spacing. The alternating array is periodic with primitive period $2L$. A uniform electric field $E_\perp$ is applied in the transverse $y$ direction. The quantum motion is retained in the two-dimensional $(x,y)$ plane.
\vspace{0.5pc}

\noindent
For a weak transverse external field, the alternating Coulomb lattice may be viewed as a two-dimensional Bloch system~\cite{Bloch1929} with a longitudinal periodic coordinate and a confined transverse degree of freedom. The transverse electric field induces a controlled force along the confined transverse direction. Expansion about a lattice site yields the restoring stiffness \(\kappa\propto\zeta(3)/L^3\), which fixes the oscillator length \(a_y\), while the field couples neighboring oscillator states through the \(y\)-dependent matrix element. In the two-dimensional representation, the Coulomb interaction depends on the relative displacement \(\sqrt{(x-nL)^2+y^2}\) and enters the Bloch secular equation through the matrix element \(\langle r,\nu|V_C(x,y)|s,\mu\rangle\). Evaluation of this matrix element in the reciprocal representation introduces the dimensionless parameter \(\eta_{r-s}\), which combines the longitudinal reciprocal vector with the transverse oscillator length \(a_y\). Because \(\eta_{r-s}\) appears in the argument of the Coulomb matrix function, its small- and large-argument limits distinguish two asymptotic coupling regimes: for small \(\eta_{r-s}\), the interaction acquires a logarithmic dependence with parity-selective transverse-state couplings, whereas for large \(\eta_{r-s}\) it approaches the corresponding one-dimensional Coulomb-chain limit. The strict one-dimensional Coulomb reduction is further extended by a current-conservation-based regularization of its singular sector. The transverse oscillator length obtained from the two-dimensional lattice then supplies a natural inherited short-distance scale, leading to an additional reciprocal coupling whose difference from the large-\(\eta_{r-s}\) two-dimensional form remains explicitly parametric. The associated one-dimensional real-space interaction reproduces the alternating Madelung factor \(M_{\rm 1D}=2\ln2\)~\cite{kittel}. The two-dimensional formulation therefore has a dual role: it determines the lattice-generated transverse restoring stiffness and retains the transverse degree of freedom required for field-induced harmonic-state coupling, while recovering the corresponding one-dimensional Coulomb description asymptotically.
\vspace{0.5pc}

\noindent The corresponding finite-order Brillouin-zone spectra show that the small \(\eta_{r-s}\) regime produces a stronger coupling condition with enhanced branch separation, while the large \(\eta_{r-s}\) spectrum approaches the Kronig--Penney~\cite{KronigPenney1931} reference toward the band-edge limit, while retaining distinct curvature within the Brillouin zone~\cite{AshcroftMermin}.
\vspace{0.5pc}

\noindent
The manuscript is organized as follows. Section~\ref{sec:sec2} introduces the alternating Coulomb lattice, the transverse electric field, and Bloch periodicity along the lattice axis. Section 3 develops the transverse harmonic basis and Coulomb matrix elements, examines the small- and large- \(\eta_{r-s}\) limits, and then connects the large \(\eta_{r-s}\) regime to the corresponding one-dimensional Coulomb chain and its regularization.  Section~\ref{sec:sec4} presents the first-Brillouin-zone reduction, the finite-order Coulomb spectra across representative \(\eta_{r-s}\) regimes, and a comparison with the Kronig--Penney reference. The principal results are summarized in Section~\ref{sec:sec5}.
\section{Alternating Coulomb lattice and Bloch formulation}
\label{sec:sec2}
We consider an infinite alternating-charge array embedded in the two-dimensional
$(x,y)$ plane,
\begin{equation}
	x_n=nL,
	\qquad
	q_n=(-1)^n q,
	\qquad
	n\in\mathbb Z,
\end{equation}
where \(L\) is the nearest-neighbor spacing. The primitive period of the alternating
lattice is therefore \(2L\). A weak uniform electric field \(E_\perp\) is applied along
the transverse \(y\) direction.
\vspace{0.5pc}

\noindent
The corresponding two-dimensional Coulomb interaction is
\begin{equation}
	V_C(x,y)
	=
	\frac{q^2}{4\pi\varepsilon}
	\sum_{n\in\mathbb Z}
	\frac{(-1)^n}
	{\sqrt{(x-nL)^2+y^2}},
\end{equation}
where \(\varepsilon\) denotes the effective permittivity. The transverse electric-field
contribution may be written locally as
\begin{equation}
	V_E^{(n)}(y)
	=
	-q_nE_\perp y
	=
	-(-1)^n qE_\perp y.
\end{equation}
The real-space single-particle Hamiltonian is then
\begin{equation}
	H(x,y)
	=
	-\frac{\hbar^2}{2m}
	\left(
	\frac{\partial^2}{\partial x^2}
	+
	\frac{\partial^2}{\partial y^2}
	\right)
	+
	V_C(x,y)
	+
	V_E(x,y),
\end{equation}
and the stationary problem is
\begin{equation}
	H(x,y)\Psi_k(x,y)
	=
	E(k)\Psi_k(x,y).
\end{equation}
Since the translational periodicity is confined to the lattice direction, we define
\begin{equation}
	G=\frac{\pi}{L},
	\qquad
	G_r=rG,
	\qquad
	r\in\mathbb Z,
\end{equation}
and impose the Bloch condition
\begin{equation}
	\Psi_k(x+2L,y)
	=
	e^{i2kL}\Psi_k(x,y).
\end{equation}
The transverse coordinate remains non-periodic and is retained as an independent
degree of freedom.
\vspace{0.5pc}

\noindent
Accordingly, the Bloch state may be written in the reciprocal form
\begin{equation}
	\Psi_k(x,y)
	=
	\sum_{r\in\mathbb Z}
	e^{i(k+G_r)x}\,
	\chi_r(y),
	\label{eq:Bloch_unspecified_y}
\end{equation}
where the transverse functions \(\chi_r(y)\) have not yet been specified. Their
structure is determined below by extracting the local transverse response of the
Coulomb lattice about a periodic site.

\section{Two-dimensional Bloch formulation and Coulomb reductions}
\label{sec:sec3}
The transverse functions \(\chi_r(y)\) in Eq.~\eqref{eq:Bloch_unspecified_y} are
constructed from the local response of the alternating Coulomb lattice about a
periodic site. To isolate the transverse degree of freedom, we choose a reference
lattice node and expand the Coulomb separation
\[
\sqrt{(x-nL)^2+y^2}
\]
for small transverse displacement \(y\), while excluding the self-site contribution.
At a lattice node the relative longitudinal separation reduces to an integer multiple
of \(L\), so that the leading \(y\)-dependent term produces a harmonic restoring
potential. This nodal reduction defines the transverse stiffness, oscillator length,
and harmonic basis used in the full two-dimensional Bloch expansion.

\subsection{Harmonic response and transverse basis}
At the reference lattice node, excluding the self-site contribution, the relative longitudinal separation is an integer multiple of \(L\). For small transverse displacement \(y\),

\begin{equation}
	\frac{1}{\sqrt{(nL)^2+y^2}}
	=
	\frac{1}{|n|L}
	-
	\frac{y^2}{2|n|^3L^3}
	+
	O(y^4),
	\qquad n\neq 0 .
\end{equation}
The quadratic contribution defines the transverse restoring potential

\begin{equation}
	V_{\rm lat}^{(2)}(y)
	=
	\frac12 \kappa y^2 ,
\end{equation}
with

\begin{equation}
	\kappa
	=
	-\frac{q^2}{4\pi\varepsilon L^3}
	\sum_{n\neq0}
	\frac{(-1)^n}{|n|^3}.
\end{equation}
Using the Dirichlet eta-function representation of the Riemann zeta function~\cite{DLMFzeta},

\begin{equation}
	\sum_{n=1}^{\infty}
	\frac{(-1)^{n-1}}{n^3}
	=
	\eta(3)
	=
	\left(1-2^{-2}\right)\zeta(3)
	=
	\frac34\zeta(3),
\end{equation}
the restoring stiffness becomes

\begin{equation}
	\kappa
	=
	\frac{3q^2\zeta(3)}
	{8\pi\varepsilon L^3}.
\end{equation}
The transverse field then shifts the local equilibrium position according to
\begin{equation}
	V_n(y)
	=
	\frac12\kappa y^2-q_nE_\perp y
	=
	\frac12\kappa
	\left(y-y_n^{(0)}\right)^2
	-
	\frac{q^2E_\perp^2}{2\kappa},
\end{equation}
where

\begin{equation}
	y_n^{(0)}
	=
	\frac{q_nE_\perp}{\kappa}
	=
	(-1)^n\frac{qE_\perp}{\kappa}.
\end{equation}
The corresponding oscillator frequency and characteristic transverse length are

\begin{equation}
	\omega_y
	=
	\sqrt{\frac{\kappa}{m}},
	\qquad
	a_y
	=
	\sqrt{\frac{\hbar}{m\omega_y}}
	=
	\left(
	\frac{\hbar^2}{m\kappa}
	\right)^{1/4}.
\end{equation}
For a weak static transverse field, the displacement may be incorporated through
shifted oscillator functions or treated perturbatively through the linear field matrix
elements. For brevity, and to retain the neighboring-state selection rule explicitly,
we use the unshifted harmonic basis

\begin{equation}
	\phi_\nu(y)
	=
	\frac{1}
	{\pi^{1/4}\sqrt{2^\nu\nu!a_y}}
	H_\nu\!\left(\frac{y}{a_y}\right)
	\exp\!\left(-\frac{y^2}{2a_y^2}\right),
\end{equation}
with

\begin{equation}
	\epsilon_\nu^{(y)}
	=
	\hbar\omega_y
	\left(
	\nu+\frac12
	\right).
\end{equation}
The  transverse function \(\chi_r(y)\) is therefore expanded in the harmonic basis as 
\begin{equation}
	\chi_r(y)=\sum_\nu C_{r\nu}(k)\phi_\nu(y).
\end{equation}
\subsection{Product basis and two-dimensional Hamiltonian}

The Bloch state is expanded as
\begin{equation}
	\Psi_k(x,y)
	=
	\sum_{r\in\mathbb{Z}}
	\sum_{\nu=0}^{\infty}
	C_{r\nu}(k)\,
	e^{i(k+G_r)x}\phi_\nu(y).
	\label{eq:Psi2D}
\end{equation}
Projection of the stationary Schr\"{o}dinger equation onto this product basis gives
\begin{equation}
	\sum_{s,\mu}
	H_{r\nu,s\mu}^{(2D)}(k)
	C_{s\mu}(k)
	=
	E(k)C_{r\nu}(k).
	\label{eq:secular2D}
\end{equation}
In this product basis, the two-dimensional Hamiltonian matrix is
\begin{align}
	H_{r\nu,s\mu}^{(2D)}(k)
	&=
	\left[
	\frac{\hbar^2(k+G_r)^2}{2m}
	+
	\epsilon_\nu^{(y)}
	\right]
	\delta_{rs}\delta_{\nu\mu}
	+
	V_{r\nu,s\mu}^{({\rm C})}
	+
	E_{r\nu,s\mu}^{(\perp)}.
	\label{eq:H2Dmatrix}
\end{align}
The transverse-field term is
\begin{equation}
	E_{r\nu,s\mu}^{(\perp)}
	=
	-qE_\perp\,\delta_{rs}\,
	\langle\nu|y|\mu\rangle.
	\label{eq:Eperp}
\end{equation}
Since
\begin{equation}
	y
	=
	\frac{a_y}{\sqrt{2}}
	\left(a+a^\dagger\right),
\end{equation}
one obtains
\begin{equation}
	E_{r\nu,s\mu}^{(\perp)}
	=
	-\frac{qE_\perp a_y}{\sqrt{2}}\,
	\delta_{rs}
	\left[
	\sqrt{\mu+1}\,\delta_{\nu,\mu+1}
	+
	\sqrt{\mu}\,\delta_{\nu,\mu-1}
	\right].
	\label{eq:Eperp-selection}
\end{equation}
Thus a static transverse field has no first-order diagonal oscillator matrix element in this basis and enters through the neighboring channels \(\mu\rightarrow\mu\pm1\). The finite-basis energies are obtained from
\begin{equation}
	\det\!\left[
	H^{(2D)}(k)-EI
	\right]
	=
	0.
	\label{eq:det2D}
\end{equation}
\subsection{Coulomb matrix element and the coupling parameter
	\texorpdfstring{$\eta_{r-s}$}{eta}}

The Coulomb contribution to the two-dimensional Hamiltonian is

\begin{equation}
	V^{(C)}_{r\nu,s\mu}
	=
	\langle r,\nu|V_C|s,\mu\rangle .
\end{equation}
Using the product basis, the longitudinal part of the bra and ket gives

\begin{equation}
	e^{-i(k+G_r)x}e^{i(k+G_s)x}
	=
	e^{-i(G_r-G_s)x}
	=
	e^{-iG_{r-s}x},
	\qquad
	G_{r-s}=G_r-G_s .
\end{equation}
The reciprocal transfer \(G_{r-s}\) therefore arises directly from the
Bloch-state matrix element. To obtain the corresponding Coulomb coefficient,
we write the alternating lattice charge distribution as

\begin{equation}
	\rho(x)
	=
	q\sum_{n\in\mathbb Z}(-1)^n\delta(x-nL).
\end{equation}
Since the primitive cell has length \(2L\), its reciprocal coefficients are

\begin{equation}
	\rho_l
	=
	\frac{q}{2L}
	\left[1-e^{-iG_lL}\right]
	=
	\frac{q}{2L}
	\left[1-(-1)^l\right]
	=
	\frac{q}{L}g_l ,
\end{equation}
where

\begin{equation}
	g_l=\frac{1-(-1)^l}{2}.
\end{equation}
Thus only odd reciprocal components of the alternating lattice survive.
For the matrix element connecting the reciprocal states \(r\) and \(s\),
the relevant component is \(l=r-s\), so that

\begin{equation}
	g_{r-s}
	=
	\frac{1-(-1)^{r-s}}{2}.
\end{equation}
The longitudinal transform of the two-dimensional Coulomb factor can be expressed in terms of the modified Bessel function of the second kind, \(K_0\) (Macdonald function)~\cite{Arfken}, through the identity
\begin{equation}
	\int_{-\infty}^{\infty}
	\frac{e^{-iG_{r-s}x}}
	{\sqrt{x^2+y^2}}\,dx
	=
	2K_0\!\left(|G_{r-s}||y|\right).
\end{equation}
Combining the alternating-lattice coefficient with this transform gives the
reciprocal Coulomb component

\begin{equation}
	V^{(C)}_{r-s}(y)
	=
	\mathcal V_{r-s}
	K_0\!\left(|G_{r-s}||y|\right),
\end{equation}
with

\begin{equation}
	\mathcal V_{r-s}
	=
	\frac{q^2}{2\pi\varepsilon L}\,
	g_{r-s}.
\end{equation}
Projection onto the transverse harmonic states then yields

\begin{equation}
	V^{(C)}_{r\nu,s\mu}
	=
	\mathcal V_{r-s}
	\int_{-\infty}^{\infty}
	\phi_\nu(y)
	K_0\!\left(|G_{r-s}||y|\right)
	\phi_\mu(y)\,dy .
	\label{eq:K0matrix}
\end{equation}

\subsection{Small \(\eta_{r-s}\) logarithmic regime of the Coulomb term}

For \(\eta_{r-s}\ll1\), the region carrying the dominant oscillator weight samples the logarithmic form
\begin{equation}
    K_0(z)
    =
    -\ln\!\left(\frac{z}{2}\right)
    -
    \gamma
    +
    O(z^2\ln z),
    \qquad
    z\ll1,
    \label{eq:K0-small}
\end{equation}
where \(\gamma\) is Euler's constant. Equation~\eqref{eq:K0matrix} then gives
\begin{equation}
    V_{r\nu,s\mu}^{({\rm SR})}
    \simeq
    -\mathcal V_{r-s}
    \Bigg\{
    \left[
        \ln\!\left(
            \frac{\eta_{r-s}}{2}
        \right)
        +
        \gamma
    \right]\delta_{\nu\mu}
    +
    \mathcal L_{\nu\mu}
    \Bigg\},
    \label{eq:VSR}
\end{equation}
where
\begin{equation}
    \mathcal L_{\nu\mu}
    = \int_{-\infty}^{\infty}
    \phi_\nu(y)
    \ln\!\left(\frac{|y|}{a_y}\right)
    \phi_\mu(y)\,dy
    =\int_{-\infty}^{\infty}
    \Phi_\nu(u)\ln|u|\Phi_\mu(u)\,du.
    \label{eq:Lnumu}
\end{equation}
We use the dimensionless form, with 
\( u = y/a_y\) and \(\Phi_\nu(u) = \sqrt{a_y}\,\phi_\nu(y)\).
The Hermite polynomial can be represented as 
\begin{equation}
    H_n(\xi)
    =
    n!
    \sum_{j=0}^{\lfloor n/2\rfloor}
    \frac{(-1)^j(2\xi)^{n-2j}}
    {j!(n-2j)!},
\end{equation}
the logarithmic cross term may be reduced to a finite Gamma-function sum. For \(\nu+\mu\) even,
\begin{align}
    \mathcal L_{\nu\mu}
    &=
    \frac{\nu!\mu!}
    {\sqrt{\pi\,2^{\nu+\mu}\nu!\mu!}}
    \sum_{j=0}^{\lfloor\nu/2\rfloor}
    \sum_{k=0}^{\lfloor\mu/2\rfloor}
    \frac{(-1)^{j+k}
    2^{\nu+\mu-2j-2k}}
    {j!k!(\nu-2j)!(\mu-2k)!}
    \nonumber\\
    &\qquad\times
    \frac{1}{2}
    \Gamma'\!\left(
        \frac{\nu+\mu-2j-2k+1}{2}
    \right),
    \label{eq:GammaCross}
\end{align}
while parity gives
\begin{equation}
    \mathcal L_{\nu\mu}=0,
    \qquad
    \nu+\mu\ {\rm odd}.
\end{equation}
The small \(\eta_{r-s}\) sector therefore retains explicit transverse-state structure and same-parity Coulomb mixing even when \(E_\perp=0\).

\subsection{Large \(\eta_{r-s}\) regime}

For \(\eta_{r-s} \gtrsim 1\), the Coulomb matrix element is retained in its
oscillator-weighted form,
\begin{equation}
	V_{r\nu,s\mu}^{({\rm LR},2D)}
	=
	\mathcal V_{r-s}
	\mathcal I_{\nu\mu}(\eta_{r-s}),
\end{equation}
where
\begin{equation}
	\mathcal I_{\nu\mu}(\eta_{r-s})
	=
	\int_{-\infty}^{\infty}
	\Phi_\nu(u)
	K_0(\eta_{r-s} |u|)
	\Phi_\mu(u)\,du .
\end{equation}
Thus the transverse oscillator basis remains explicitly coupled through
the \(K_0\) kernel.
\vspace{0.5pc}

\noindent For large \(\eta_{r-s}\), the \(K_0\) kernel becomes increasingly localized near the transverse origin \(u=0\), so that the Coulomb matrix element is dominated by the behavior of the oscillator functions near the center. Introducing \(t=\eta_{r-s}u\) in \(\mathcal I_{\nu\mu}\) gives, to leading order for states with nonzero amplitude at the origin,
\begin{equation}
	 \mathcal I_{\nu\mu}(\eta_{r-s}) \sim \frac{\pi\Phi_\nu(0)\Phi_\mu(0)} {\eta_{r-s}}.
\end{equation}

\noindent Since \(\eta_{r-s}=|G_{r-s}|a_y\), the resulting Coulomb matrix acquires the reciprocal hierarchy
\begin{equation}
	V^{(C)}_{r\nu,s\mu}\propto\frac{1}{|G_{r-s}|}\propto\frac{1}{|r-s|},
	\label{eq:Vclr}
\end{equation}
 which provides the asymptotic connection to the one-dimensional Coulomb-chain form developed in the following subsection.
\vspace{0.5pc}

\subsection{One-dimensional Coulomb chain and reciprocal Hamiltonian}
The one-dimensional periodic interaction is given by
\begin{equation}
	V_{\rm 1D}(x)
	=
	V_0
	\sum_{p=1}^{\infty}
	g_{p}
	\frac{\cos(G_{p} x)}p,
\end{equation}
where
\begin{equation}
	g_{p}
	=
	\frac{1-(-1)^{p}}{2}
\end{equation}
retains the odd reciprocal components of the alternating lattice.
\vspace{0.5pc}

\noindent
For the strictly one-dimensional alternating chain, the Coulomb interaction has the reciprocal form obtained from the Fourier components of the alternating \(1/|x|\) potential. In the Bloch representation, the off-diagonal coupling therefore carries the factor \(g_{r-s}/|r-s|\) with \(p=|r-s|\), giving the same reciprocal hierarchy obtained asymptotically in Eq~\eqref{eq:Vclr}.
\vspace{0.5pc}

\noindent
The reciprocal Hamiltonian becomes
\begin{equation}
	H_{rs}^{({\rm 1D})}(k)
	=
	\frac{\hbar^2(k+G_r)^2}{2m}\delta_{rs}
	+
	(1-\delta_{rs})
	g_{|r-s|}
	\frac{V_0}{2|r-s|}.
\end{equation}
The Coulomb interaction also gives the familiar one-dimensional
alternating Madelung factor in real space,
\begin{equation}
	M_{\rm 1D}=2\ln2,
\end{equation}
so that the real-space Coulomb sum and the reciprocal
\(1/|r-s|\) hierarchy provide equivalent representations of the same
long-range interaction.
\vspace{0.5pc}

\noindent
The distinction between a strict one-dimensional Coulomb chain and the large \(\eta_{r-s}\) representation of the two-dimensional version lies in the treatment of the transverse electric
field. In the two-dimensional formulation, \(y\) remains an independent
degree of freedom and the field enters explicitly through the transverse
oscillator matrix elements. 

\subsection{Regularized one-dimensional Coulomb extension}
The strictly one-dimensional Coulomb chain nevertheless remains a singular-potential problem, so the behavior near the lattice sites requires a regularization prescription for a physical realization.
\vspace{0.5pc}

\noindent
A regularization of this type may be introduced through the kinematic
current-conservation condition. In the Ermakov formulation, the corresponding
closure generates an additional quantum contribution proportional to
\(\hbar^2/(2m)\), with the admissible singular forms determined by the
auxiliary solution and its initial conditions~\cite{IQJFregularization}.
The associated class contains inverse-square and related trigonometric,
hyperbolic, and elliptic singular structures. Here we retain only the
inverse-square limit and use it to construct a regularized one-dimensional
extension.
\vspace{0.5pc}

\noindent
The preceding two-dimensional formulation already provides the characteristic
oscillator length
\begin{equation}
	a_y
	=
	\left(
	\frac{\hbar^2}{m\kappa}
	\right)^{1/4},
\end{equation}
through the lattice-generated transverse stiffness. In the reduced
one-dimensional problem, \(a_y\) is therefore used as an inherited
short-distance regularization scale rather than introducing an independent
cutoff parameter.
\vspace{0.5pc}

\noindent
For a lattice node \(x_n=nL\), the singular inverse-square term is taken in the
form
\begin{equation}
	V_{\rm sing}(x)
	=
	\frac{\hbar^2}{8m}
	\frac{1}{(x-x_n)^2}.
\end{equation}
A symmetric displacement of the poles away from the real axis,
\begin{equation}
	x-x_n
	\;\longrightarrow\;
	(x-x_n)\pm i a_y,
\end{equation}
leads to the regularized real-axis form
\begin{equation}
	V_{\rm reg}(x)
	=
	\frac{\hbar^2}{8m}
	\sum_{n\in\mathbb Z}
	\frac{1}{(x-nL)^2+a_y^2}.
	\label{eq:Vreg1D}
\end{equation}

\noindent
The contribution of Eq.~\eqref{eq:Vreg1D} to the Bloch Hamiltonian is obtained
from the reciprocal matrix element
\begin{equation}
	\langle r|V_{\rm reg}|s\rangle
	=
	\frac{\hbar^2}{8m}
	\left\langle r\left|
	\sum_{n\in\mathbb Z}
	\frac{1}{(x-nL)^2+a_y^2}
	\right|s\right\rangle .
\end{equation}
For Bloch factors \(e^{i(k+G_r)x}\), the quasimomentum \(k\) cancels between
the bra and ket, leaving the reciprocal transfer
\begin{equation}
	e^{-i(k+G_r)x}e^{i(k+G_s)x}
	=
	e^{-iG_{r-s}x}.
\end{equation}
The required kernel follows from the identity
\begin{equation}
	\int_{-\infty}^{\infty}
	\frac{e^{-iG_{r-s}x}}
	{x^2+a_y^2}\,dx
	=
	\frac{\pi}{a_y}
	e^{-a_y|G_{r-s}|}.
\end{equation}
Since the regularizing lattice has period \(L\), whereas the
alternating-charge Bloch cell has period \(2L\), only even reciprocal
transfers contribute. With
\begin{equation}
	p=|r-s|,
	\qquad
	G_p=pG,
\end{equation}
the regularization matrix element becomes
\begin{equation}
	\langle r|V_{\rm reg}|s\rangle
	=
	\begin{cases}
		R_p, & p=0,2,4,\ldots,\\[4pt]
		0,   & p=1,3,5,\ldots,
	\end{cases}
	\qquad
	p=|r-s|.
\end{equation}
where
\begin{equation}
	R_p
	=
	\frac{\hbar^2\pi}{8m a_y L}
	e^{-a_y|G_p|}.
\end{equation}
\vspace{0.5pc}

\noindent
The regularized reciprocal Hamiltonian consequently becomes
\begin{equation}
	H_{rs}^{({\rm REG})}(k)
	=
	H_{rs}^{\rm 1D}(k)+ \langle r|V_{reg}|s\rangle .
	\label{eq:Hreg1D}
\end{equation}
Thus the original alternating Coulomb hierarchy occupies the odd reciprocal
sector, while the inverse-square regularization introduces an additional
even-transfer contribution.
\vspace{0.5pc}

\noindent
For the first-order truncation \(r=-1,0,+1\), the regularized Hamiltonian is
\begin{equation}
	H_{\rm REG}^{(3)}(k)
	=
	\begin{pmatrix}
		\alpha(k-G)^2+R_0
		&
		U
		&
		R_2
		\\
		U
		&
		\alpha k^2+R_0
		&
		U
		\\
		R_2
		&
		U
		&
		\alpha(k+G)^2+R_0
	\end{pmatrix},
	\label{eq:Hreg3}
\end{equation}
where
\begin{equation}
	\alpha=\frac{\hbar^2}{2m},
	\qquad
	U=\frac{V_0}{2}.
\end{equation}
\vspace{0.5pc}

\noindent
The regularized one-dimensional Bloch branches follow from
\begin{equation}
	\det
	\left[
	H_{\rm REG}^{(3)}(k)-EI
	\right]
	=0.
	\label{eq:secularReg}
\end{equation}
In this form, the singular one-dimensional Coulomb chain is extended without
introducing a new phenomenological length scale: the regularization length is
inherited from the lattice-generated transverse stiffness of the preceding
two-dimensional construction. The resulting even-transfer term provides a
direct route for examining short-distance corrections to the one-dimensional
Bloch spectrum.
\vspace{0.5pc}

\noindent
For comparison with the representative large \(\eta_{\rm LR}\) two-dimensional reduction used in the Brillouin-zone analysis below, the regularized one-dimensional Hamiltonian may be written relative to the corresponding long-range matrix as

$$ H^{({\rm REG})}(k)-H^{({\rm LR})}(k;\eta_{\rm LR}) = E_G\left(\frac14-\frac{1}{\eta_{\rm LR}^2}\right)I + R_2 \begin{pmatrix} 0&0&1\\ 0&0&0\\ 1&0&0 \end{pmatrix}, $$
\vspace{0.5pc}

\noindent
for the representative regularization choice \(\eta_{\rm reg}=1\). The first term is a purely parametric difference between the inherited regularization offset \(R_0=E_G/4\) and the transverse energy scale \(E_G/\eta_{\rm LR}^2\), while the second isolates the additional even-transfer coupling introduced by regularization. At \(\eta_{\rm LR}=2\) the diagonal mismatch vanishes, leaving only the finite \(R_2\) correction; this special value reflects the chosen energy scales rather than a general equivalence of the two descriptions.

\section{Brillouin-zone reduction and finite-order comparison}
\label{sec:sec4}

Because translational periodicity is along $x$, the Bloch momentum lies in the first Brillouin zone
\begin{equation}
	-\frac{G}{2}
	\leq
	k
	\leq
	\frac{G}{2},
	\qquad
	G=\frac{\pi}{L}.
	\label{eq:BZ}
\end{equation}
The two-dimensional Coulomb problem and the one-dimensional Coulomb chain share the same one-dimensional Bloch interval. In the long-range sector, their reciprocal interactions display the same \(1/|r-s|\) hierarchy, while only the two-dimensional formulation retains the transverse oscillator degree of freedom.
\vspace{0.5pc}

\noindent
For a three-component reciprocal truncation,
\begin{equation}
	r=-1,0,+1,
\end{equation}
the one-dimensional Hamiltonian is
\begin{equation}
	H_3^{(1D)}(k)
	=
	\begin{pmatrix}
		T_{-1} & U_{-1,0} & U_{-1,+1}\\
		U_{0,-1} & T_0 & U_{0,+1}\\
		U_{+1,-1} & U_{+1,0} & T_{+1}
	\end{pmatrix},
\end{equation}
where
\begin{equation}
	T_r
	=
	\frac{\hbar^2(k+G_r)^2}{2m},
\end{equation}
and
\begin{equation}
	U_{rs}
	=
	g_{r-s}
	\frac{V_0}{2|r-s|},
	\qquad
	r\neq s.
\end{equation}
\vspace{0.5pc}

\noindent
For a fixed transverse oscillator state $\nu$, the corresponding projected two-dimensional matrix is
\begin{equation}
	H_{3,\nu}^{(2D)}(k)
	=
	\begin{pmatrix}
		T_{-1}+\epsilon_\nu^{(y)} & W_{-1,0}^{(\nu)} & W_{-1,+1}^{(\nu)}\\
		W_{0,-1}^{(\nu)} & T_0+\epsilon_\nu^{(y)} & W_{0,+1}^{(\nu)}\\
		W_{+1,-1}^{(\nu)} & W_{+1,0}^{(\nu)} & T_{+1}+\epsilon_\nu^{(y)}
	\end{pmatrix},
\end{equation}
with
\begin{equation}
	W_{rs}^{(\nu)}
	=
	\mathcal V_{r-s}\,
	\mathcal I_{\nu\nu}
	\!\left(\eta_{r-s}\right).
	\label{eq:WetaBZ}
\end{equation}
For a fixed transverse state \(\nu\), the term \(\epsilon_\nu^{(y)}\) provides a common diagonal energy scale, while the reciprocal Coulomb elements determine the branch mixing. Diagonalization of the finite-order matrix gives three Bloch branches,

\begin{equation}
	E_j(k)
	=
	\operatorname{eig}_j\!\left[
	H_{3,\nu}^{(2D)}(k)
	\right],
	\qquad
	j=-,0,+,
\end{equation}
which are compared below with the corresponding one-dimensional reference spectrum.
\subsection{First Brillouin-zone spectra}

\begin{figure}[!htbp]
	\centering
	\includegraphics[
	width=0.9\textwidth,
	height=0.9\textheight,
	keepaspectratio
	]{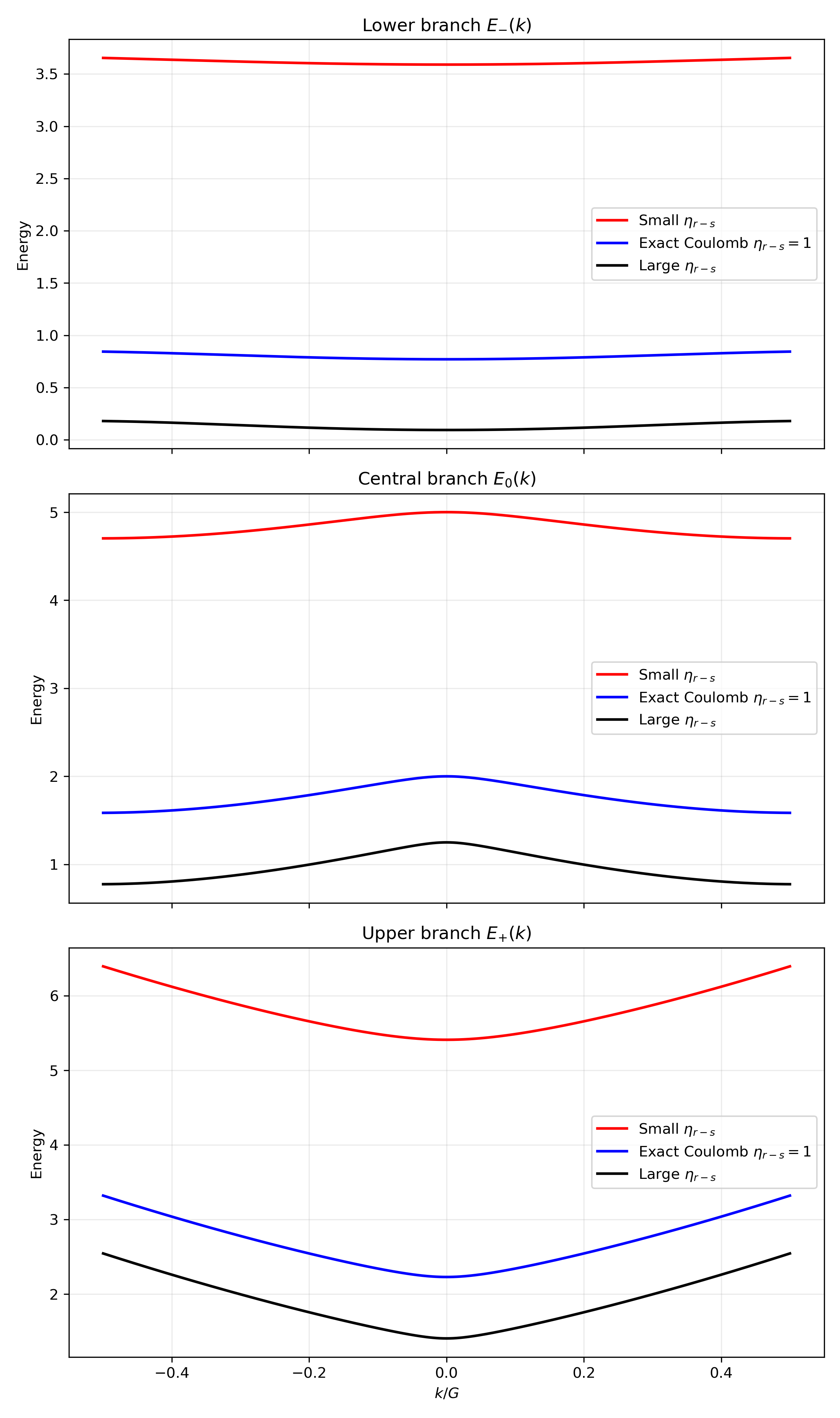}
	\caption{Branch-resolved Bloch energy dispersions \(E_-(k)\), \(E_0(k)\), and
		\(E_+(k)\) over the first Brillouin zone for the two-dimensional Coulomb lattice
		in the small \(\eta_{r-s}\), exact-Coulomb \(\eta_{r-s}=1\), and
		large \(\eta_{r-s}\) regimes, evaluated at \(E_\perp=0\).}
	\label{fig:etaBZ}
\end{figure}

\begin{figure}[!htbp]
	\centering
	\includegraphics[
	width=0.9\textwidth,
	height=0.9\textheight,
	keepaspectratio
	]{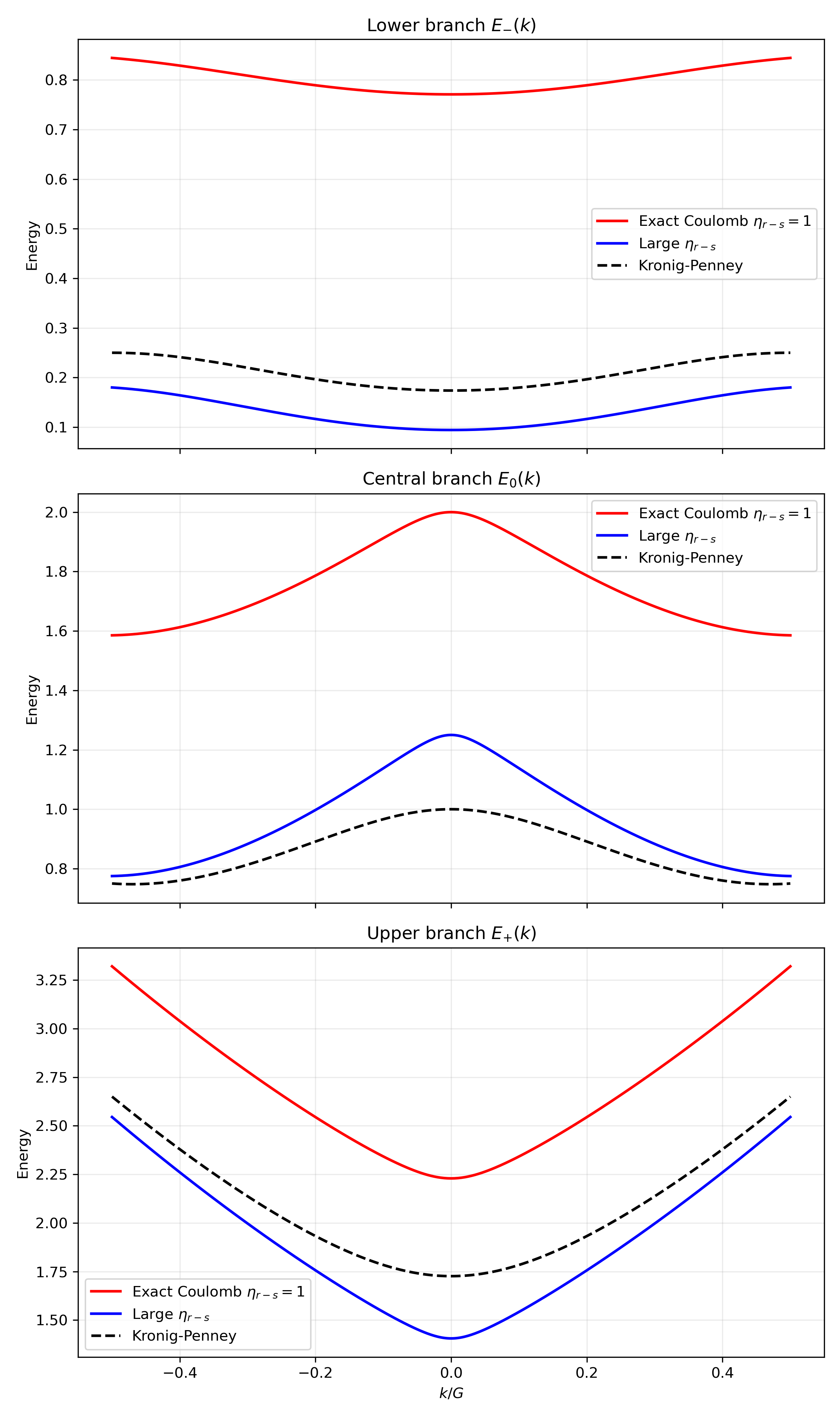}
	\caption{Branch-resolved Bloch energy dispersions \(E_-(k)\), \(E_0(k)\), and
		\(E_+(k)\) over the first Brillouin zone for the exact-Coulomb
		\(\eta_{r-s}=1\) case, the large \(\eta_{r-s}\) Coulomb regime, and the
		Kronig--Penney benchmark, evaluated at \(E_\perp=0\).}
	\label{fig:etaKGPLRBZ}
\end{figure}

To illustrate the effect of Coulomb coupling strength on the Bloch spectrum, we
first compare three representative interaction regimes of the same
two-dimensional Coulomb construction: a small \(\eta_{r-s}\) regime, an
intermediate exact-Coulomb case at \(\eta_{r-s}=1\), and a large \(\eta_{r-s}\)
regime. The corresponding branch-resolved Coulomb dispersions are shown in Fig.~\ref{fig:etaBZ}. As a secondary benchmark, the exact-Coulomb \(\eta_{r-s}=1\) and large \(\eta_{r-s}\) spectra are compared with the Kronig--Penney reference in Fig.~\ref{fig:etaKGPLRBZ}.
\vspace{0.5pc}

\noindent
The numerical comparison is expressed in units of the longitudinal kinetic
scale
\begin{equation}
	E_G=\alpha G^2,
	\qquad
	\alpha=\frac{\hbar^2}{2m},
\end{equation}
and we use the illustrative values
\begin{equation}
	G=1,
	\qquad
	\alpha=1,
	\qquad
	L=\frac{\pi}{G}=\pi,
\end{equation}
so that \(E_G=1\). The Coulomb scale is taken as
\begin{equation}
	U=0.30E_G,
	\qquad
	V_0=2U=0.60E_G.
\end{equation}
For the small \(\eta_{r-s}\) regime we choose
\begin{equation}
	\eta_{\rm SR}=0.5,
	\qquad
	\epsilon_{y,{\rm SR}}^{(0)}
	=
	\frac{E_G}{\eta_{\rm SR}^2}
	=
	4E_G.
\end{equation}
In the fixed-\(\nu=0\) approximation, the off-diagonal logarithmic coupling is

\begin{equation}
	W_1^{({\rm SR})}
	=
	-\mathcal V_1
	\left[
	\ln\!\left(\frac{\eta_{\rm SR}}{4}\right)
	+
	\frac{\gamma}{2}
	\right],
	\qquad
	\mathcal V_1=\frac{V_0}{2},
\end{equation}
which gives
\begin{equation}
	W_1^{({\rm SR})}\simeq0.537E_G.
\end{equation}
The corresponding \(3\times3\) reciprocal Hamiltonian is

\begin{equation}
	H^{({\rm SR})}(k)
	=
	\begin{pmatrix}
		\alpha(k-G)^2+\epsilon_{y,{\rm SR}}^{(0)} & W_1^{({\rm SR})} & 0 \\
		W_1^{({\rm SR})} & \alpha k^2+\epsilon_{y,{\rm SR}}^{(0)} & W_1^{({\rm SR})} \\
		0 & W_1^{({\rm SR})} & \alpha(k+G)^2+\epsilon_{y,{\rm SR}}^{(0)}
	\end{pmatrix}.
\end{equation}
To display the intermediate regime explicitly, we also include an exact-Coulomb (EC)
case at
\begin{equation}
	\eta_{\rm EC}=1,
	\qquad
	\epsilon_{y,{\rm EC}}^{(0)}
	=
	\frac{E_G}{\eta_{\rm EC}^2}
	=
	E_G.
\end{equation}
The fixed-\(\nu=0\) Coulomb matrix element is obtained from the full
oscillator-weighted kernel,
\begin{equation}
	\mathcal{I}_{00}(1)
	=
	\int_{-\infty}^{\infty}
	\Phi_0(u)\,
	K_0(|u|)\,
	\Phi_0(u)\,du,
\end{equation}
so that
\begin{equation}
	W_1^{({\rm EC})}
	=
	\mathcal V_1 \mathcal{I}_{00}(1).
\end{equation}
Numerically,
\begin{equation}
	\mathcal{I}_{00}(1)\simeq1.2509,
	\qquad
	W_1^{({\rm EC})}\simeq0.3753E_G.
\end{equation}
The corresponding finite-order matrix is
\begin{equation}
	H^{({\rm EC})}(k)
	=
	\begin{pmatrix}
		\alpha(k-G)^2+\epsilon_{y,{\rm EC}}^{(0)} & W_1^{({\rm EC})} & 0 \\
		W_1^{({\rm EC})} & \alpha k^2+\epsilon_{y,{\rm EC}}^{(0)} & W_1^{({\rm EC})} \\
		0 & W_1^{({\rm EC})} & \alpha(k+G)^2+\epsilon_{y,{\rm EC}}^{(0)}
	\end{pmatrix}.
\end{equation}
For the large \(\eta_{r-s}\) regime we take

\begin{equation}
	\eta_{\rm LR}=2,
	\qquad
	\epsilon_{y,{\rm LR}}^{(0)}
	=
	\frac{E_G}{\eta_{\rm LR}^2}
	=
	0.25E_G.
\end{equation}
In this sector the reciprocal coupling follows the projected long-range
hierarchy,
\begin{equation}
	H_{rs}^{({\rm LR})}(k)
	=
	\left[
	\alpha(k+rG)^2+\epsilon_{y,{\rm LR}}^{(0)}
	\right]\delta_{rs}
	+
	(1-\delta_{rs})
	g_{|r-s|}
	\frac{V_0}{2|r-s|}.
\end{equation}
Under the first-order truncation \(r=-1,0,+1\), this reduces to
\begin{equation}
	H^{({\rm LR})}(k)
	=
	\begin{pmatrix}
		\alpha(k-G)^2+\epsilon_{y,{\rm LR}}^{(0)} & U & 0 \\
		U & \alpha k^2+\epsilon_{y,{\rm LR}}^{(0)} & U \\
		0 & U & \alpha(k+G)^2+\epsilon_{y,{\rm LR}}^{(0)}
	\end{pmatrix}.
\end{equation}
The lower, central, and upper Bloch branches are obtained in each case from the
three eigenvalues
\begin{equation}
	E_j^{(\lambda)}(k)
	=
	\mathrm{eig}_j\!\left[H^{(\lambda)}(k)\right],
	\qquad
	j=-,0,+,
\end{equation}
with \(\lambda={\rm SR},{\rm EC},{\rm LR}\). Fig.~\ref{fig:etaBZ} shows that
the small \(\eta_{r-s}\) regime produces the strongest branch separation, the
exact-Coulomb \(\eta_{r-s}=1\) case provides an intermediate spectrum, and the
large \(\eta_{r-s}\) regime approaches the weaker-coupling end of the same
Coulomb family.

\FloatBarrier
\vspace{0.5pc}

\noindent
For comparison with a standard one-dimensional periodic model, we also evaluate
the \(3\times3\) Kronig--Penney Hamiltonian
\begin{equation}
	H^{({\rm KP})}(k)
	=
	\begin{pmatrix}
		\alpha(k-G)^2+V_K & -V_K & V_K \\
		-V_K & \alpha k^2+V_K & -V_K \\
		V_K & -V_K & \alpha(k+G)^2+V_K
	\end{pmatrix},
\end{equation}
where \(V_K\) denotes the Kronig–Penney coupling scale, with \(V_K=0.30E_G\).
\vspace{0.5pc}

\noindent
Its three branches
\begin{equation}
	E_j^{({\rm KP})}(k)=\mathrm{eig}_j\!\left[H^{({\rm KP})}(k)\right]
\end{equation}
 are compared in Fig.~\ref{fig:etaKGPLRBZ} with both the exact-Coulomb \(\eta_{r-s}=1\) spectrum and the large \(\eta_{r-s}\) Coulomb limit. The comparison provides a conventional periodic reference while retaining the \(\eta_{r-s}\) dependent Coulomb spectra as the primary result.
\vspace{0.5pc}

\noindent
For a weak but finite transverse field, the small \(\eta_{r-s}\)
two-dimensional model acquires additional off-diagonal couplings between
neighboring transverse states. For sufficiently small field strengths, these
terms act perturbatively and do not qualitatively alter the branch structure of
the zero-field spectra shown here.
\section{Summary and Conclusions}
\label{sec:sec5}
An infinite alternating-charge array under a weak transverse electric field has been formulated as a two-dimensional Bloch problem with longitudinal periodicity and a confined transverse degree of freedom. Expansion about a lattice site gives the transverse restoring stiffness
\begin{equation}
	\kappa
	=
	\frac{3q^2\zeta(3)}
	{8\pi\varepsilon L^3},
\end{equation}
which defines the oscillator scale \(a_y\). The transverse field couples neighboring oscillator states through \(E_{r\nu,s\mu}^{(\perp)}\), while the Coulomb matrix elements are controlled by the dimensionless parameter
\begin{equation}
	\eta_{r-s}= |G_{r-s}|a_y .
\end{equation}
For \(\eta_{r-s}\ll1\), the logarithmic Coulomb form produces parity-dependent transverse-state couplings, whereas for larger \(\eta_{r-s}\) the full oscillator-weighted interaction is retained. Its long-range reduction gives the same \(1/|r-s|\) reciprocal hierarchy as the corresponding one-dimensional Coulomb chain, whose real-space form yields the alternating Madelung factor \(M_{\rm 1D}=2\ln2\).
\vspace{0.5pc}

\noindent
The branch-resolved spectra show that the small \(\eta_{r-s}\) regime corresponds to the stronger Coulomb-coupling condition, with enhanced separation of the lower, central, and upper branches. In the large \(\eta_{r-s}\) regime, the Coulomb spectrum approaches the Kronig--Penney benchmark toward the band-edge limit, while retaining distinct curvature around the center of the Brillouin zone as a consequence of the two-dimensional Coulomb construction.
\vspace{0.5pc}

\noindent
The strict one-dimensional Coulomb reduction is further extended by a current-conservation-based regularization of its singular sector. The transverse oscillator length obtained from the two-dimensional lattice then supplies a natural inherited short-distance scale, leading to an additional reciprocal coupling whose difference from the large \(\eta_{r-s}\) two-dimensional form remains explicitly parametric.
\vspace{0.5pc}

\noindent
\textbf{Declaration of Interests.}
The research is an independent work of the author and declares no conflicts of interest with the employer, IBM Research, Albany, NY, USA.
\vspace{0.5pc}

\noindent\textbf{Use of Generative AI.}
Generative AI tools were used in the preparation of this manuscript solely for language editing, grammar refinement, and equation cross-verification. These tools were not used to generate the scientific results, derivations, or core interpretations presented here. All mathematical checks, technical conclusions, and final editorial decisions remain the sole responsibility of the author.

\newpage

\FloatBarrier

\section*{Glossary of Symbols}
\addcontentsline{toc}{section}{Glossary of Symbols}
For convenience, the principal symbols used throughout the manuscript are
summarized below.
\begin{center}
	\begin{tabular}{p{0.22\textwidth} p{0.7\textwidth}}
		\hline
		\textbf{Symbol} & \textbf{Definition} \\
		\hline
		
		\(L\) &
		Nearest-neighbor spacing of the alternating lattice.
		\\[0.3em]
		
		\(2L\) &
		Primitive period of the alternating-charge array.
		\\[0.3em]
		
		\(q_n\) &
		Charge at lattice site \(n\).
		\\[0.3em]
		
		\(E_\perp\) &
		Uniform electric field applied along the transverse direction.
		\\[0.3em]
		
		\(G_{r-s}\) &
		Reciprocal transfer between Bloch components \(r\) and \(s\).
		\\[0.3em]
			
		\(p\) &
		Magnitude of the reciprocal-index transfer, \(p =|r-s|\).
		\\[0.3em]
		
		\(g_{p}\) &
		Alternating-lattice selector that retains odd reciprocal transfers.
		\\[0.3em]
		
		\(\Psi_k(x,y)\) &
		Two-dimensional Bloch wave function.
		\\[0.3em]
		
		\(\chi_r(y)\) &
		Transverse amplitude associated with reciprocal component \(r\).
		\\[0.3em]
		
		\(\kappa\) &
		Transverse restoring stiffness generated by the alternating Coulomb lattice.
		\\[0.3em]
		
		\(\omega_y\) &
		Transverse harmonic-oscillator frequency.
		\\[0.3em]
		
		\(a_y\) &
		Characteristic transverse oscillator length.
		\\[0.3em]
		
		\(\phi_\nu(y)\) &
		Normalized transverse harmonic-oscillator eigenfunction.
		\\[0.3em]
		
		\(\nu,\mu\) &
		Transverse harmonic-oscillator quantum numbers.
		\\[0.3em]
		
		\(\epsilon_\nu^{(y)}\) &
		Energy of the transverse harmonic-oscillator state.
		\\[0.3em]
		
		\(C_{r\nu}(k)\) &
		Expansion coefficient of the two-dimensional Bloch state.
		\\[0.3em]
		
		\(H_{r\nu,s\mu}^{(2D)}(k)\) &
		Matrix element of the two-dimensional Bloch Hamiltonian.
		\\[0.3em]
		
		\(V_{r\nu,s\mu}^{(C)}\) &
		Coulomb matrix element between reciprocal and transverse basis states.
		\\[0.3em]
		
		\(\mathcal V_{r-s}\) &
		Reciprocal Coulomb prefactor associated with transfer \(r-s\).
		\\[0.3em]
		
		\(K_0\) &
		Modified Bessel function of the second kind of order zero
		(Macdonald function).
		\\[0.3em]
		
		\(\eta_{r-s}\) &
		Dimensionless parameter coupling the reciprocal transfer to the transverse oscillator length.
		\\[0.3em]
		
		\(\mathcal I_{\nu\mu}(\eta)\) &
		Oscillator-weighted Coulomb integral.
		\\[0.3em]
		
		\(\mathcal L_{\nu\mu}\) &
		Logarithmic transverse-state matrix element in the small-\(\eta\) regime.
		\\[0.3em]
		
		\(E_G\) &
		Longitudinal kinetic-energy scale used in the finite-order simulations.
		\\[0.3em]
		
		\(E_-(k),\,E_0(k),\,E_+(k)\) &
		Lower, central, and upper eigenvalue branches of the finite-order Bloch matrix.
		\\[0.3em]
		
		\(M_{\rm 1D}\) &
		One-dimensional alternating Madelung factor, (\(2\ln2\)).
		\\[0.3em]
		
		\(V_{reg}(x)\) &
		Regularized one-dimensional Coulomb potential contribution.
		\\
		
		\hline
	\end{tabular}
\end{center}

\end{document}